\documentclass[reprint,aps,prb,superscriptaddress,english,nolongbibliography,twocolumn]{revtex4-2}

\usepackage{graphicx,dcolumn,braket}

\usepackage[table,xcdraw,dvipsnames]{xcolor}
\usepackage[colorlinks=true,citecolor=blue,urlcolor=blue,linkcolor=blue]{hyperref}

\usepackage{amsmath,amssymb,bm,mathrsfs,amsfonts}

\begin{document}

%\title{Principal component analysis of localization transitions in disordered quantum systems}
\title{An analysis of localization transitions using non-parametric unsupervised learning}

\author{Carlo Vanoni}
\email{cvanoni@sissa.it}
%\thanks{These authors contributed equally}
\affiliation{SISSA -- International School for Advanced Studies, via Bonomea 265, 34136, Trieste, Italy}
\affiliation{INFN Sezione di Trieste, Via Valerio 2, 34127 Trieste, Italy}
\author{Vittorio Vitale}
\email{vittorio.vitale@lpmmc.cnrs.fr}
%\thanks{These authors contributed equally}
\affiliation{Univ. Grenoble Alpes, CNRS, LPMMC, 38000 Grenoble, France}
\date{\today}

\begin{abstract}
    We propose a new viewpoint on the study of localization transitions in disordered quantum systems, showing how critical properties can be seen also as a geometric transition in the data space generated by the classically encoded configurations of the disordered quantum system. We showcase our approach to the Anderson model on regular random graphs, known for displaying features of interacting systems, despite being a single-particle problem.
    We estimate the transition point and critical exponents in agreement with the best-known results in the literature. We provide a simple and coherent explanation of our findings, discussing the applicability of the method in real-world scenarios with a modest number of measurements.
\end{abstract}

\maketitle

\section{Introduction}
In the last decades, a huge effort has been devoted to understanding non-equilibrium phases of matter, which circumvent the maximum-entropy constraint of thermal equilibrium~\cite{Rigol2008Thermalization,Polkovnikov2011Colloquium}. Within this class of problems, the complete characterization of the breakdown of ergodicity induced by disorder in quantum systems represents one of the standing open quests~\cite{Abanin2019colloquium}.
Together with the huge theoretical effort, there has been increasingly large attention to these unusual phases of matter also in the experimental community; as a consequence of the possibility of realizing theoretical models in the laboratory~\cite{smith2016many,schreiber2015observation,shtanko2023uncovering,white2020observation,billy2008direct,roati2008anderson}. 
However, it is often difficult to find observables that are readily accessible and theoretically predictable.
\begin{figure}
    \centering
    \includegraphics[width=\linewidth]{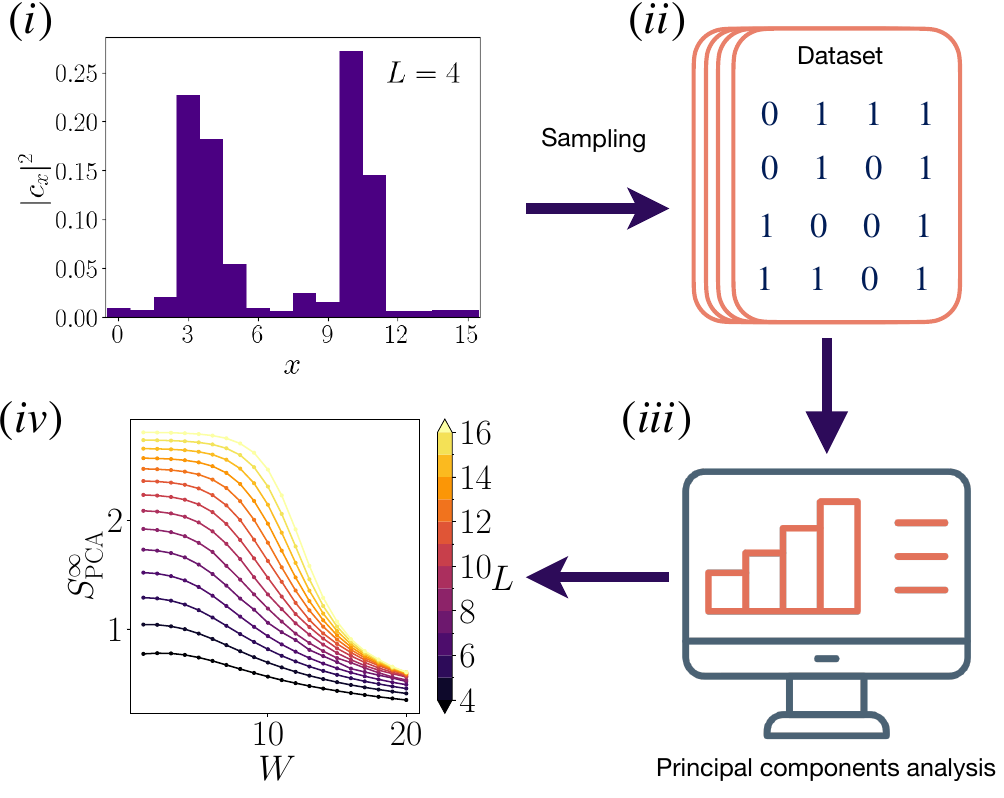}
    \caption{Sketch of the approach used in this work. Given a quantum state $\ket{\psi}$, $(i)$ we chose a basis $\ket{\psi_x}$ and we sample (measure) the state according to the probability distribution $|c_x|^2=|\langle \psi_x | \psi \rangle|^2$. In this example, we consider a system with Hilbert space dimension equal to $2^L$ ($L=4$). $(ii)$ We encode the measurement outcome as a string of zeros and ones corresponding to the binary representation of the integer $x\in [0,2^L]$, labeling the basis state. $(iii)$ We perform the principal components analysis (PCA) and extract the information we are interested in by averaging over several realizations of the disorder. In\,$(iv)$, we show the behavior of $S^{\infty}_\textrm{PCA}=-\ln \lambda_1$ (see Eq.~\ref{eq:renyiPCA} and text for details) as a function of the disorder strength $W$, for several sizes $L$.}
    \label{fig:fig1}
\end{figure}

In this work, we propose a data-science-inspired method in the context of disordered quantum systems, and, in particular, we show that localization transitions can also be investigated through the behavior of the classically encoded configurations in data space. To this end, we employ principal components analysis (PCA), which is used to detect the most relevant directions in data space and to compress (to project) the data set toward the significant and restricted manifold~\cite{wold1987principal,jolliffe2002principal}.
From the eigendecomposition of the sample covariance matrix, we introduce the  R\'enyi-entropy of the normalized eigenvalues $\lambda_j$'s ($\sum_{j=1}^d \lambda_j=1$) as
\begin{equation}\label{eq:renyiPCA}
S^{(n)}_\mathrm{PCA}:=\frac{1}{1-n} \ln \sum_{j=1}^d \lambda^{n}_{j},
\end{equation}
and we show analytically that $S_\mathrm{PCA}$, i.e. $S^{(n=1)}_\mathrm{PCA}$, is linked to the participation entropy, often employed for investigating disordered induced transitions~\cite{sierant2023universality,vanoni2023renormalization,kutlin2024investigating}.
Therefore, unlike usual non-parametric approaches, our physically informed method is guaranteed to work, in the limit of sufficiently large samples.
We show that the infinite-order $S^{\infty}_\mathrm{PCA}$ can be employed to estimate the critical point with remarkable accuracy in agreement with recent results~\cite{sierant2023universality,suntajs2023similarity} and displays universal behavior around the transition.
Moreover, we employ a data set whose dimension is smaller than the full Hilbert space, thus being readily applicable in modern quantum simulators where large data sets of snapshots of the state of the system are routinely collected~\cite{Kunkel2019,Brydges2019,hoke2023quantum,vitale2023estimation}.

We remark here that data-science-inspired approaches have already found several successful applications in various fields, ranging from classical and quantum statistical physics~\cite{Wetzel2017,TiagoPRX2021,TiagoPRXQ2021,Turkeshi2022data,panda2023nonparametric,vitale2023topological,verdel2023datadriven,mendessantos2023wave} to molecular science and quantum chemistry~\cite{mehta2019high,carrasquilla2020machine}. 

To prove the validity of our approach, we showcase it on a prototypical example of disordered quantum systems displaying a localization transition: the Anderson model on random regular graphs (RRGs). 
The latter displays Anderson localization~\cite{Anderson1958absence} with an usual scaling of expectation values with system size ~\cite{tikhonov2021AndersonMBL,tikhonov2016Anderson,kravtsov2018non,de2014anderson,Parisi_2019,de2020subdiffusion,bera2018return,sierant2023universality,vanoni2023renormalization}, and is especially hard to tackle numerically~\cite{Pino2020Scaling,sierant2023universality}, thus being the ideal test bench for the method we propose.
In the Supplementary Material~\cite{SupplMat} we provide more details on $S_{\textrm{PCA}}$ and our analysis, and present results for a many-body disordered model that is believed to display a localization transition, showing that the method presented in this work is effective also for interacting systems.

%It is well known that, in non-interacting quantum systems, the presence of impurities can lead to the absence of transport and thus to localization of the wavefunction describing the quantum degrees of freedom, a phenomenon known as Anderson localization~\cite{Anderson1958absence}. More recently, it has been proposed that such a disorder-induced localization can be robust against the presence of interactions and lead to many-body localization (MBL)~\cite{Basko06,Pal10,Nandkishore15,Imbrie17,ros2015integrals}, on which the debate is still open~\cite{DeRoeck2017Stability,Luitz2017How,Thiery2018manybody,Morningstar2022Avalanches,suntajs2023similarity,Sierant2022Challenges}.
%The Anderson model on random regular graphs (RRGs) displays a MBL-like unusual scaling of expectation values with system size ~\cite{tikhonov2021AndersonMBL,tikhonov2016Anderson,kravtsov2018non,de2014anderson,Parisi_2019,de2020subdiffusion,bera2018return,sierant2023universality,vanoni2023renormalization}, and thus it is especially hard to tackle numerically~\cite{Pino2020Scaling,sierant2023universality} and the ideal test bench for the method we propose.

The remainder of the work is structured as follows. 
In Sec.~\ref{sec:WFSPCA} we present the method we use to sample the wavefunction and the rationale behind the analysis of the data. In Sec.~\ref{sec:M&R} we exploit the Anderson model on RRGs, giving a quick presentation of the system and its properties, and showcasing the effectiveness of the approach. Finally, in Sec.~\ref{sec:concl} we give our conclusions and discuss possible outlooks.

\section{Wavefunction sampling and analysis}\label{sec:WFSPCA}
The interest in non-parametric unsupervised learning methods relies on their vast range of applicability, a consequence of their agnosticism towards the problem under analysis. Such versatility is ensured by the fact that the only required input is a data set, which in principle can come from any sort of source, and whose geometrical properties are analyzed to extract information from the underlying physical system. 
In our case, the data sets consist of matrices in which each row corresponds to a single snapshot of a wave function, i.e. a measurement in a given basis (see Fig.~\ref{fig:fig1} $(ii)$). However, the method presented here may be applied to a plethora of experimental and numerical situations.

In practice, let us assume to have a state described by
\begin{equation}\label{eq:genericstate}
\ket{\psi}=\sum_{x=1}^{\mathcal{N}} c_x \ket{\psi_x},
\end{equation}
where $\{\ket{\psi_x}\}_{x=1,\dots,\mathcal{N}}$ is a suitable basis in the Hilbert space $\mathcal{H}$ of dimension $\mathcal{N}=\mathrm{dim}(\mathcal{H})$.
The sampling of $\ket{\psi}$ amounts to sample, according to the probabilities $|c_x|^2$, the corresponding basis vectors $\ket{\psi_x}$.
The choice of the relevant basis and the encoding of the sampling into an actual data set is one of the aspects to be investigated.
For example, considering a chain of qubits, one could measure a state $\ket{\psi}$ in the computational basis and getting as an outcome a string of zeros and ones.
In this work, we label as $X_i= ( n_{i,1},\dots,n_{i,d} )$ an element of the configuration space, where each $n_{i,x}$, called `feature', encodes some information of the sampled state; e.g. in the previous scenario, each feature corresponds to the measured state of the qubit (say $0$ or $1$) and the total number of features $d$ is equal to the size of the system.
The full target data set is a collection of $N_r$ repetitions of $X_i$ :
\begin{equation}\label{eq:data sets}
    X=\left( X_1,X_2,\dots ,X_{N_r} \right)
\end{equation}
and can be represented as a ($N_r \times d$) matrix $X_{i,j}$.

Concretely, the method we employ is the following. We define the centered data set $X_c$, whose elements are 
\begin{equation}
\label{eq:center}
    (X_c)_{i,j} = X_{i,j}-\frac{1}{N_r}\sum_i X_{i,j}
\end{equation} and compute the covariance matrix $C= X_c^T X_c/(N_r-1)$.  
Then, we perform the eigendecomposition $C = V^T K V$, where $K=\mathrm{diag}(k_1,\dots,k_r)$ is the diagonal matrix of the $r$ eigenvalues of $C$ ordered in descending order, and $V=(v_1,\dots,v_r)$ is the rotation whose columns $v_j$ identify the $j$-th relevant directions. In the new reference frame defined by $V$, the variance of the data along the $j$-th direction is given by $k_j$, and thus $\lambda_j \equiv k_j/(\sum_i k_i)$ represents the percentage of encoded information along the direction $v_j$ and is dubbed \emph{$j$-th explained variance ratio} ($\lambda_1 > \lambda_2 > \dots > \lambda_r$). 

The motivation for our study comes from the understanding that the 
$S_\mathrm{PCA}$ --- recently introduced as a measure of the information content of a physical data set~\cite{panda2023nonparametric,verdel2023datadriven} --- is connected to the participation entropy. This is particularly relevant since the participation entropy is the typical quantity of interest when studying disordered systems and is used for estimating an order parameter: the fractal dimension~\cite{sierant2023universality,vanoni2023renormalization,kravtsov2018non,kutlin2023anatomy,Sierant2022Universal}.
The presence of such a connection between $S_{\textrm{PCA}}$ and participation entropy is intriguing as, in usual scenarios, when non-parametric estimators are employed, a clear physical picture is missing. Here we show that studying the principal components is physically meaningful as they are connected to an order parameter and thus they are guaranteed to store information of the physical process.

Let us link $S_{\textrm{PCA}}$ and participation entropy by considering the sampling of a state written as in Eq.\eqref{eq:genericstate}.
For each sample on the basis $\{\ket{\psi_x}\}_{x=1,\dots,\mathcal{N}}$, we obtain as an outcome an integer $x$ with probability $|c_x|^2$. Let us assume to encode this as an $\mathcal{N}$-dimensional vector with only a non-zero entry corresponding to the index $x$ of the sampled basis vector $\ket{\psi_x}$. Then, the element of the configuration space would be vectors of the type $X_i = (0, \dots, 0, 1, 0, \dots, 0)$.
In Ref.~\cite{SupplMat}, we prove that for a large enough number of samplings $N_r \gg \mathcal{N}$, one gets $C =X^T X/(N_r-1)=  \mathrm{diag}(|c_1|^2,|c_2|^2,\dots,|c_{\mathcal{N}}|^2)$ and the $S_{PCA}$ becomes
\begin{equation}\label{eq:SPCAparticipationentropy}
    S_{\mathrm{PCA}}=-\sum_j |c_j|^2 \ln |c_j|^2,
\end{equation}
which is exactly the definition of the participation entropy.
However, let us observe that the correspondence we have shown is only true in the limiting case $N_r \gg \mathcal{N}$ and that the data set contains exponentially large vectors.
Therefore, one could ask if working with different choices of encoding and at finite sampling could provide estimates on the critical parameters of the transition as well.

We show that this is valid by studying the behavior of $S^{\infty}_{\textrm{PCA}}=-\ln{\lambda_1}$. 
The rationale behind this is that $\lambda_1$ contains all the information needed for spotting the localization transition. In fact, in the localized phase we expect a single wavefunction coefficient $c_x$ to be dominant. The sampled data set should be such that the first explained variance ratio $\lambda_1$ is much larger than all the others, namely there should be a single predominant direction in the data space manifold.
On the other hand, in the ergodic regime, all wavefunction components should be of the same order, and thus the principal components of the samplings should have all the same importance.  There should not be a preferred direction in data space, and the explained variance ratios should vanish with the system size (since the normalization $\sum_j \lambda_j =1$ is enforced).

In the remainder, we showcase these predictions by exploiting the Anderson model on RRGs. We find that with an appropriate analysis, it is possible to retrieve remarkably good estimations on the position of the critical point of the disordered induced transition and perform a clean finite size scaling. We do so by employing a modest number of measurements and obtain results that are in agreement with the literature and with statistical errors that are compatible with state-of-the-art methods.

\begin{figure*}
    \includegraphics[width=\textwidth]{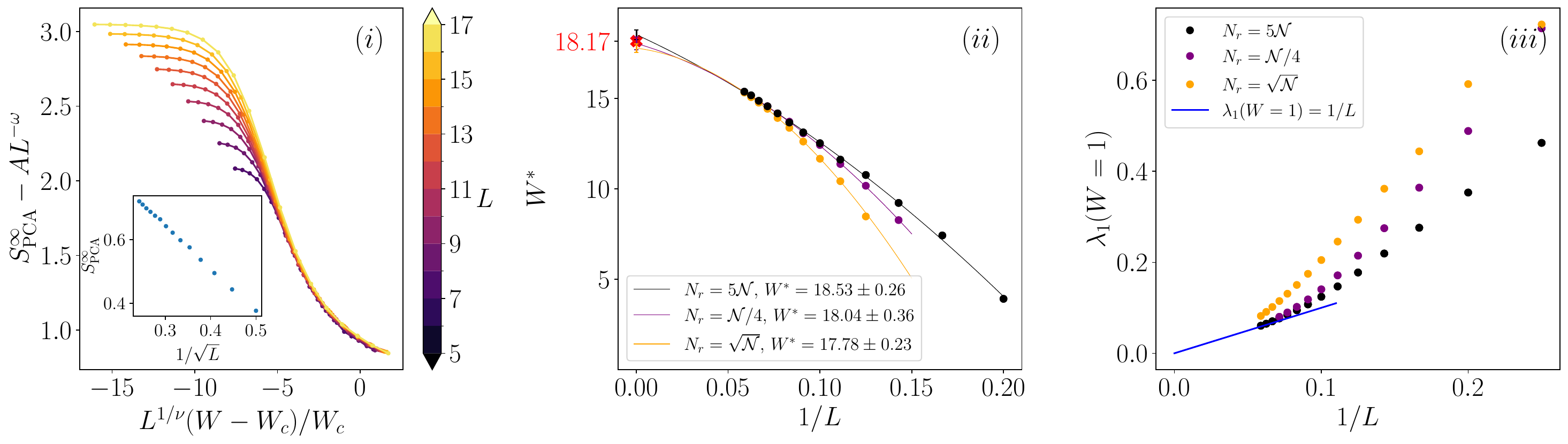}
    \caption{$(i)$ Plot of the finite size scaling of $S^{\infty}_{\mathrm{PCA}}$ (as in Eq.~\eqref{eq:scaling}). We fix $\nu = 1$ and $\omega=1/2$ and only tune the parameter $A$ to obtain the collapse. The plot of $S^{\infty}_{\mathrm{PCA}} = - \log \lambda_1$ is reported in panel $(iv)$ of Fig.~\ref{fig:fig1}. $(i \mathrm{\,-\, inset})$ Behavior of $S_{\mathrm{PCA}}^{\infty}$ at $W=W_c$ as a function of $L^{-1/2}$; we observe $S^{\infty}_\mathrm{PCA}\sim L^{-\omega}$ with $\omega = 1/2$.
    $(ii)$ Plot of $W^*$ vs $1/L$. The extrapolation to $L\to \infty$ gives the correct position of the critical point $W_c=18.17$, denoted with a red cross. Three different sets of points are shown: black dots obtained by sampling the eigenstates $N_r = 5 \mathcal{N}$ times; purple points employing $N_r = \mathcal{N} / 4$ samples; and orange dots using $N_r = \sqrt{\mathcal{N}}$. A parabolic fit in $1/L$, the easiest curve accounting for the curvature of the points, is performed and the critical value of $W$ is extrapolated at $1/L = 0$. The fitting functions are $W^* = 18.53 \pm 0.26 + (-46.73\pm 4.66)/L + (-128.27 \pm 18.57)/L^2$ (black), $W^* = 18.04 \pm 0.36 + (-29.31\pm 7.01)/L + (-272.44 \pm 32.85)/L^2$ (purple) and $W^* = 17.78 \pm 0.23 + (-13.45\pm 5.40)/L + (-475.69 \pm 31.19)/L^2$ (orange). In Ref.~\cite{SupplMat} we elaborate more on the fitting procedure. $(iii)$ Behavior of $\lambda(W=1)$ vs $1/L$, both for $N_r = 5 \mathcal{N}$ (black) and $N_r = \mathcal{N}/4$ (orange) samples. It is expected, at large sizes, that $\lambda_1$ in the ergodic phase goes as $1/L$, being the inverse of the rank of the matrix $C$. This behavior is indeed reached at large sizes as all sets of points approach the $1/L$ line in blue, which is a guide for the eye.}
    \label{fig:fig2}
\end{figure*}
\section{Model and results}
\label{sec:M&R}
Let us consider the Anderson model for a single quantum particle on a random regular graph (RRG). The Hamiltonian of the model is~\cite{Anderson1958absence}
\begin{equation}\label{eq:HamiltonianAnderson}
    H = - \sum_{\langle x, y \rangle} \left( \ket{x}\bra{y} + \ket{y}\bra{x} \right) + \sum_x \epsilon_x \ket{x} \bra{x},
\end{equation}
where $x,y$ are integers that label the node of the graph. The Hamiltonian consists of two terms. The first one is the adjacency matrix of the graph ($\langle x, y \rangle$ denotes nearest neighbor sites), in which, by construction, each node (or vertex) has connectivity $K_0$ (i.e. fixed vertex degree $\mathcal{D}=K_0+1$). The second term represents a random field applied on each site, with the parameters $\epsilon_i$ being independent and identically distributed random variables sampled according to the box distribution $g(\epsilon) = \theta(|\epsilon|-W/2)/W$. Denoting with $\mathcal{N}$ the number of vertices of the graph, we introduce a length scale $L = \ln_{K_0} \mathcal{N}$, representing the diameter of the graph, i.e. the maximal length of the shortest paths connecting two nodes.

For $K_0 = 2$, which will be assumed in the rest of the paper, the critical value of the disorder is known to be $W_c \simeq 18.17$~\cite{sierant2023universality,Pino2020Scaling,Parisi_2019,Tikhonov2019RRG}.
For $W \ll W_c$ the system is ergodic, and spectral quantities in the thermodynamic limit assume the values predicted by random matrix theory. By increasing $W$ at finite system size, the model displays a crossover to the localized regime, where Poisson statistics describes the energy spectrum. Such crossover becomes a phase transition in the thermodynamic limit, with the crossover point drifting to larger $W$ as $\mathcal{N}$ is increased and reaching $W_c$ in the $\mathcal{N} \to \infty$ limit~\cite{sierant2023universality,Pino2020Scaling,vanoni2023renormalization}.
To find the critical disorder for which the whole system ceases to be ergodic, one has to focus on eigenstates near the middle of the spectrum, i.e. around zero energy for the model under consideration. This is because the eigenstates in the middle of the spectrum are those that need more disorder to localize~\cite{Mott1987MobilityEdge,van1999localization} (on the contrary, the ground state is always localized).

The numerical simulations on the model in Eq.~\eqref{eq:HamiltonianAnderson} are performed as follows. To find the eigenstates, we execute a full exact diagonalization of its matrix for $L \leq 14$, or employ the POLFED algorithm for larger system sizes~\cite{POLFED}. We calculate $\sim \sqrt{\mathcal{N}}$ eigenvectors in the middle of the spectrum. For each one (see Fig.~\ref{fig:fig1}), $(i)$ we sample, according to the probabilities $|c_x|^2$, the corresponding basis vectors $\ket{\psi_x}$. Since the problem is single-particle, we consider the basis $\ket{\psi_x}=\ket{x}$ where the particle occupies the site $x$. Then, the output of a single sampling will be the position of the particle $x$. $(ii)$ We encode the information as a $L$-dimensional vector corresponding to the binary representation of the integer $x$; $(iii)$ we perform the analysis on the data set and $(iv)$ average the results over a number of realizations of the disordered Hamiltonian in Eq.~\eqref{eq:HamiltonianAnderson} ranging from $O(10^4)$ for the smallest sizes to $O(10^2)$ for $L=17$.
%We observe that in the case of a many-body problem, the same approach could be used employing a suitable basis. For instance, taking into account a system of $L$ interacting qubits, one could sample it in the computational basis and obtain a data set comprised of $L$-dimensional strings of zeros and ones as well.

We look at the behavior of $S^{\infty}_\textrm{PCA}$ as a function of the strength of the disorder $W$ and for different sizes of the graph, that we distinguish via the length scale $L$ (see Fig.~\ref{fig:fig1}~$(iv)$).
We observe that $S^{\infty}_\textrm{PCA}=-\ln \lambda_1$ shows a crossover from the delocalized to the localized phase. In the limit of infinite disorder, the wavefunction is fully localized and it is expected that $S^{\infty}_\textrm{PCA}$ approaches $0$. On the other side, in the limiting case $W \sim 0$, there is no preferential configuration sampled. All the non-vanishing  $\lambda_j$ are the same, and thus $\lambda_1 \sim 1/L$. This holds for any $W<W_c$ in the large $L$ limit. We show the behavior of $\lambda_1$ for $W=1$ in Fig.~\ref{fig:fig2}$(iii)$, as a function of $1/L$, observing that it displays the expected behavior for large $L$.

To address the critical exponents, we perform a finite-size scaling of $S^{\infty}_{\mathrm{PCA}}$. We employ the scaling ansatz presented in Ref.~\cite{sierant2023universality} for the average gap ratio, which in our case takes the form
\begin{equation}\label{eq:scaling}
    S^{\infty}_{\mathrm{PCA}} = f((W-W_c)L^{1/\nu}) + L^{-\omega}f_1((W-W_c)L^{1/\nu}),
\end{equation}
where $f(x)$ and $f_1$ are, respectively, the leading and subleading scaling functions and $\nu$ and $\omega$ are the critical exponents. Here, $\nu$ governs the divergence of the correlation length at the critical point when $W \to W_c^{-}$ and
%, and therefore can be extracted, in principle, from any observable upon finite-size scaling and 
does not depend on the specific observable. In Ref.~\cite{sierant2023universality} it is found to be $\nu = 1$. On the other hand, $\omega$ governs the behavior of the observable under analysis at the critical point $W=W_c$. In the case of the average gap ratio, it is found $\omega = 2$~\cite{sierant2023universality}. In our case, we find $\omega = 1/2$ for $S^{\infty}_{\mathrm{PCA}}$, as it can be seen from the inset of Fig.~\ref{fig:fig2}$(i)$. Setting $\nu = 1$ we obtain a very clean collapse in Fig.~\ref{fig:fig2}$(i)$. We have approximated the subleading scaling function $f_1(x)$ with a constant $A$, which is the only free parameter of our analysis, and we have set $W_c = 18.17$~\cite{sierant2023universality,Pino2020Scaling,Parisi_2019,Tikhonov2019RRG}.

To estimate the critical point $W_c$, we study the intersection of $S^{\infty}_{\textrm{PCA}}$ with the horizontal line $S^{\infty}_{\textrm{PCA}}=1$, since the position of the intersection point $W^*$ drifts when increasing the size of the graph, approaching eventually $W_c$. Different choices of the position of the line give results compatible with the ones shown here. 
We plot the behavior of $W^*$ as a function of $1/L$ in Fig.~\ref{fig:fig2}$(ii)$. Here we report the results in the case $N_r=\sqrt{\mathcal{N}}$ (orange), $N_r=\mathcal{N}/4$ (purple) and $N_r=5\mathcal{N}$ (black) and we perform a parabolic fit in $1/L$ to estimate $W_c$. We observe that both extrapolations give a value that is compatible with the one in the literature, also in the case of a modest number of configurations sampled. In particular, we find $W_c(\sqrt{\mathcal{N}})=17.78 \pm 0.23$, $W_c(\mathcal{N}/4)=18.04 \pm 0.36$ and $W_c(5\mathcal{N})=18.53 \pm 0.26$, where the critical value of the disorder is $W_c = 18.17 \pm 0.01$.
Let us remark here that the critical value $W_c = 18.17 \pm 0.01$ is obtained by solving self-consistent equations for the propagator on the Bethe lattice~\cite{Tikhonov2019RRG}, thus allowing for a higher precision. Instead, state-of-the-art numerical methods to estimate $W_c$ on RRGs have errors on the estimates that are compatible with the ones of our approach~\cite{sierant2023universality,Pino2020Scaling}.

\section{Conclusions and outlook}
\label{sec:concl}
In this manuscript, we introduced a non-parametric unsupervised learning approach to tackle localization transitions.
We have connected analytically the eigendecomposition of the sample covariance matrix to the participation entropy, physically motivating our approach.
We have showcased it on the Anderson model on a random regular graph that, even if non-interacting, displays important features that are reminiscent of many-body localization and presents a serious challenge both analytically and numerically. 
Exploiting this example we have shown that disordered quantum systems can be characterized with data-science-inspired approaches and localization transitions can also be seen as geometric transitions in data space. 

We have studied the infinite order R\'enyi entropy $S^{\infty}_\textrm{PCA}$ of the eigenvalues covariance matrix as a function of disorder strength and system size, to extract an estimate of the critical value of the disorder $W_c$, that is remarkably in agreement with results in the literature --- in particular considering the hard challenge presented by the model investigated~\cite{Pino2020Scaling,sierant2023universality}.
As observed in Fig.~\ref{fig:fig2}$(ii)$, a modest number of measurements suffices for estimating the transition point, such that the approach described here can be considered of practical use for nowadays quantum simulators with local addressing.

Furthermore, we have performed a finite size scaling of $S^{\infty}_\textrm{PCA}$ by
employing the scaling ansatz presented in Ref.~\cite{sierant2023universality} for
the average gap ratio, and we have obtained results compatible with the literature.

We observe that the method employed requires no apriori knowledge of the physical system under investigation, being then a powerful tool also in the study of other physical scenarios, in particular many-body problems. We present results for the ‘Imbrie model'~\cite{Imbrie2016manybody,Abanin2021distinguishing}, in the Supplementary Material~\cite{SupplMat}.
The latter is believed to display many-body localization, and thus we exploit it to prove that our method is applicable also to interacting scenarios.

We note that the same analysis could be performed to tackle problems such as out-of-equilibrium phase transitions or the classification of quantum phases of matter.
Moreover, one could try to understand if this kind of approach could be used in combination with randomized measurements~\cite{elben2023randomized,cieslinski2023analysing}, to extract relevant features of many-body quantum states prepared in the laboratory.\\

\acknowledgements
The authors acknowledge Marcello Dalmonte, Cristiano Muzzi, Roberto Verdel Aranda and Antonello Scardicchio for useful discussions, and Federico Balducci, Giuliano Chiriacò, Andrea Gambassi, Piotr Sierant, and Xhek Turkeshi for comments on the manuscript. V.V. is grateful to Xhek Turkeshi for useful insights and for drawing his attention to the participation entropy. C.V. thanks Piotr Sierant for sharing the POLFED code that has been used in this work for the numerical simulations at large system sizes.
Work in Grenoble is funded by the French National
Research Agency via QUBITAF (ANR-22-PETQ-0004,
Plan France 2030).

\bibliography{references}

\newpage

\appendix

\widetext

\section{PCA entropy analysis}\label{app:SPCA}
We focus here on the PCA entropy, as described in Ref.~\cite{panda2023nonparametric}. The PCA entropy can be defined starting from the $\lambda_j$ obtained from the matrix $C$ introduced in Sec. II.
Noticing that the $\lambda_j$ satisfy (i) $\lambda_j \ge 0$ for all $j$ (as they are proportional to the squared singular values of $C$, and (ii) $\sum_j \lambda_j=1$ (by construction), we can follow Shannon's entropy formula to define  
\begin{equation}
	S_\mathrm{PCA}:=-\sum_{j=1}^d \lambda_{j} \ln (\lambda_{j}).
	\label{pca_entropy}
\end{equation}
In general, the PCA entropy in Eq.~\eqref{pca_entropy} can be used as a measure of the correlations among the input variables in the analyzed data set. Indeed, note that for an extremely `correlated' data set, which under PCA can be fully described by a single principal component (i.e., $\lambda_1\sim 1$, $\lambda_n\sim 0$, for $n\ge 2$), we get $S_\mathrm{PCA} = 0$. Instead, for a fully `uncorrelated' data set (e.g., a collection of independent random variables), for which $\lambda_j=1/d$ for all $j$, we have $S_\mathrm{PCA} = \ln \;d$.
This quantity has not been studied in quantum statistical mechanics at equilibrium so far. Here, we want to show that $S_\mathrm{PCA}$ is actually dependent on the type of encoding used for the problem at stake and draw its connection with the participation entropy. In the following, we will show that while it holds the signature of the crossover, it is not possible to efficiently extract estimates on the microscopic features of the transition (namely critical points and scaling parameters). 

\subsubsection{Encoding and participation entropy}
\begin{figure*}
\centering
    \includegraphics[width=\textwidth]{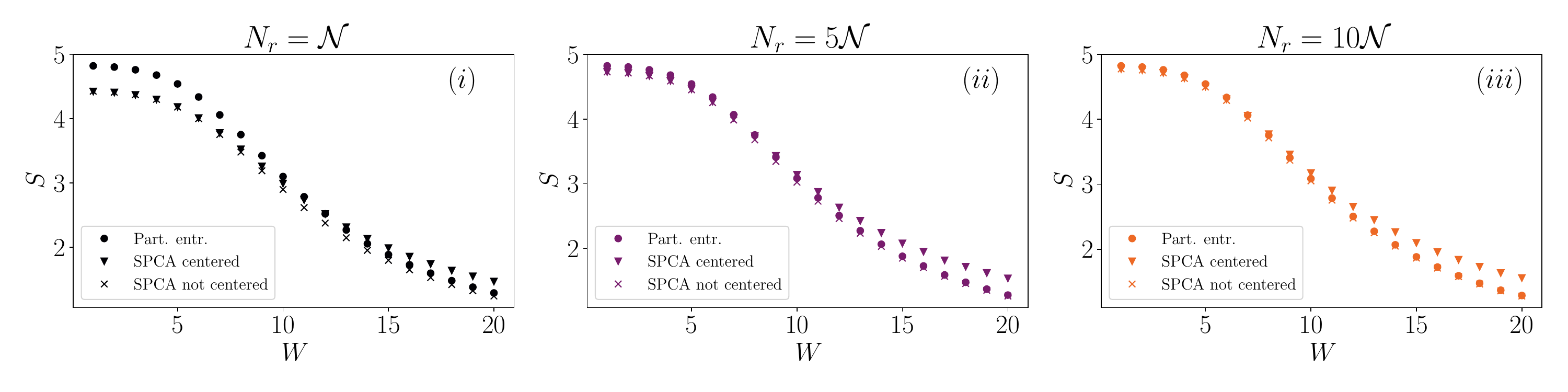}
    \caption{Comparison between participation entropy (dots), PCA entropy obtained centering the data (triangles) and PCA entropy without centering the data (crosses). We can observe that, by increasing the number of samples $N_r$ from $\mathcal{N}$ $(i)$ to $10 \mathcal{N}$ $(iii)$ the crosses converge to the dots, as expected since the non-centered PCA entropy converges to the participation entropy for $N_r \gg \mathcal{N}$. We also notice that this happens for the centered PCA entropy, but only in the ergodic regime and not in the localized phase. The motivation for this is the one presented also in the main text: at small $W$ all the eigenvalues of $C$ are of the same order and one can safely approximate the true eigenvalues with the wavefunction amplitudes, while in the localized phase this is not possible, as a single eigenvalue will become $O(1)$, invalidating the approximation.}
    \label{fig:nc_N_encod}
\end{figure*}

We started our discussion interested in studying a quantum state written as in Eq. (1) of the main text.

Let us assume to build a data set in the following way: we sample, according to the probabilities $|c_x|^2$, the corresponding basis vectors $\ket{\psi_x}$; we encode the information as a $\mathcal{N}$-dimensional vector with a single non-zero component corresponding to the index $x$. We call this `$N$-encoding'.
Each vector will be one row of the data set $X$, so that all the rows will be orthogonal to each other and the $X$ matrix can be recast in the form
\begin{equation}
\label{eq:matrix_X}
X = 
\begin{pmatrix}
    k_1 \begin{Bmatrix}
        1 & 0 & 0 &\dots & 0\\
         & & \vdots \\
        1 & 0 & 0 & \dots & 0\\
    \end{Bmatrix}\\
    k_2 \begin{Bmatrix}
        0 & 1 & 0 & \dots & 0\\
         & & \vdots \\
        0 & 1 & 0 & \dots & 0\\
    \end{Bmatrix}\\
    \vdots\\
    k_{\mathcal{N}} \begin{Bmatrix}
        0 & 0 & 0 & \dots & 1\\
         & & \vdots \\
        0 & 0 & 0 & \dots & 1\\
    \end{Bmatrix}\\
\end{pmatrix},
\end{equation}
with $\sum_{x=1}^{\mathcal{N}}k_x = N_r$.
It is immediate to observe that the sum of the columns will correspond to number of times $k_x$ a ket $\ket{\psi_x}$ has been sampled. Therefore in the limit $N_r \gg \mathrm{dim}(\mathcal{H})$ we expect $1/N_r\sum_i X_{i,j}=|c_j|^2$.
Without centering the data as in Eq. (3) of the main text (so taking $X_c = X$), one gets $X^T X = \mathrm{diag}(k_1,k_2,\dots,k_{\mathcal{N}})$ so that $C = \mathrm{diag}(|c_1|^2,|c_2|^2,\dots,|c_{\mathcal{N}}|^2)$ for $N_r \gg \mathrm{dim}(\mathcal{H})$ and the $S_{PCA}$ will assume the value
\begin{equation}\label{eq:SPCAparticipationentropy}
    S_{\mathrm{PCA}}=-\sum_j |c_j|^2 \ln |c_j|^2,
\end{equation}
which is exactly the definition of the participation entropy.
If instead one takes the centered data $(X_c)_{i,j} = X_{i,j}-\left(\sum_i X_{i,j}\right)/N_r$, it is easy to see that 
\begin{equation}
    (X_c^T X_c)_{ii} = k_i - \frac{k_i^2}{N_r}
\end{equation}
and
\begin{equation}
    (X_c^T X_c)_{ij} = - \frac{k_i k_j}{N_r^2}, \quad i \neq j.
\end{equation}
In general, the eigenvalues of $C = X_c^T X_c /(N_r - 1)$ do not coincide with $|c_i|^2$ for any choice of the $k_i$'s. However, when all the $k_i$'s are of the same order (this happens in the delocalized phase in our problem), then one can approximate $(X_c^T X_c)_{ii} \sim k_i$ and $(X_c^T X_c)_{ij} \sim 0$ and the non-centered case is retrieved, thus giving a good approximation of the participation entropy. This can be seen in Fig.~\ref{fig:nc_N_encod}.

Let us observe that the encoding proposed is not numerically efficient, for the dimension of the data set scales exponentially with the Hilbert space dimension, and also requires an exponentially large number of samples $N_r$ to recover the same information as the participation entropy. In practice, the encoding of the sampling into an actual data set and the PCA procedure can be skipped altogether as what matters is the counting of the repetitions of $\ket{\psi_x}$. Hence, in this particular scenario/encoding, the data analysis approach described here becomes pointless.

The discussion above is the reason why in this work a different encoding has been employed, let us call it `$L$-encoding'. Namely, as in the previous case we sample, according to the probabilities $|c_x|^2$, the corresponding basis vectors $\ket{\psi_x}$; we store the information as a $L$-dimensional vector that encodes the index $x$ as a binary number. Thus, each row of the data set $X$ will be a L-dimensional string of zeros and ones. Then we perform the PCA as described at the beginning of App.~\ref{app:SPCA}.

In the next section we discuss the numerical results of $S_{\textrm{PCA}}$ in the case of both encodings.

\subsubsection{Numerical simulations}
\begin{figure*}
    \includegraphics[width=\linewidth]{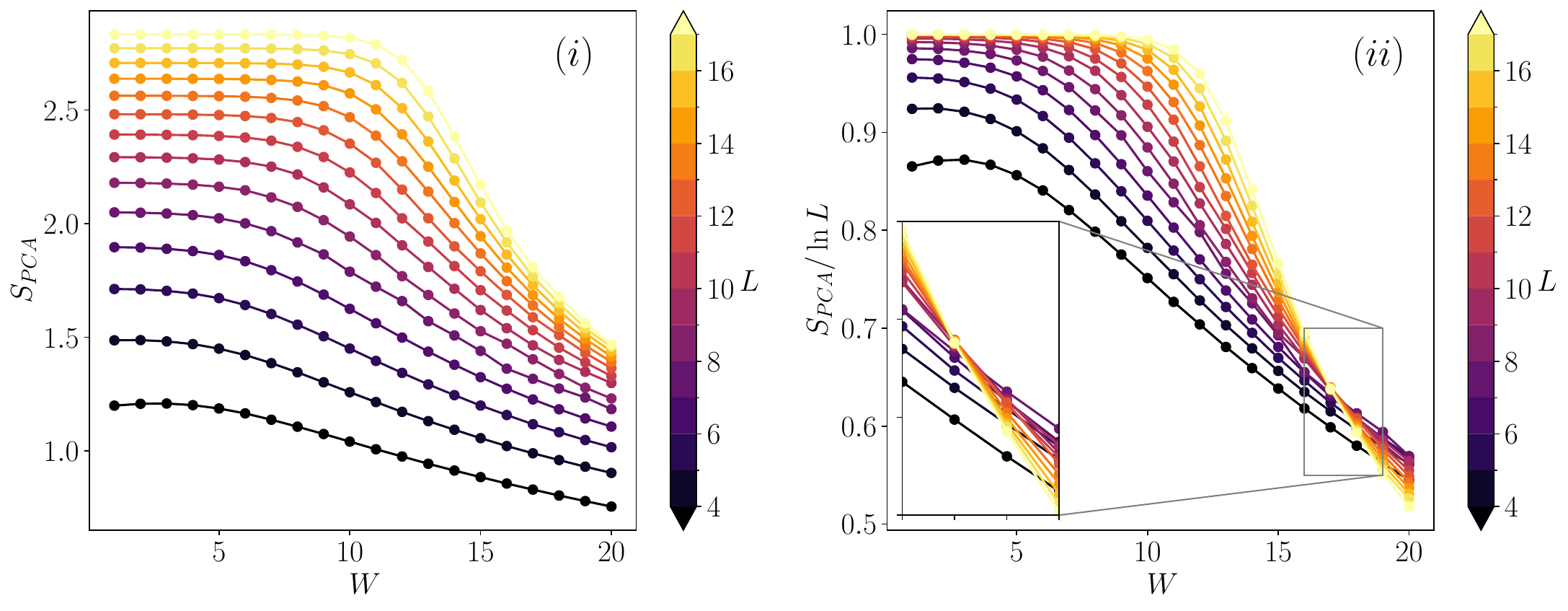}
    \caption{$(i)$ PCA entropy computed using the `$L$-encoding'. While the curves are always decreasing with $W$, as for the participation entropy, quantitatively the behavior is different. In particular, at small $W$ a long plateau develops increasing the system size, and the derivative wrt the system size does not give the participation entropy. $(ii)$ PCA entropy computed using the `$L$-encoding' and normalized with $\ln L$, so that  $SPCA/\ln L = 1$ at small $W$. As usually happens in spectral observables, e.g. the $r$-parameter or the participation entropy~\cite{sierant2023universality}, curves corresponding to different sizes cross. In our case, however, the position of the crossing point moves to smaller $W$ as $L$ grows, differently from what usually happens, forbidding to identify the correct position of the critical point.}
    \label{fig:SPCA_L}
\end{figure*}
As described in the main text, we perform an exact diagonalization of the model in Eq. (5) of the main text employing the POLFED algorithm~\cite{POLFED} and we perform a PCA on the eigenvectors in the middle of the spectrum.
\begin{figure}
    \includegraphics[width=0.5\linewidth]{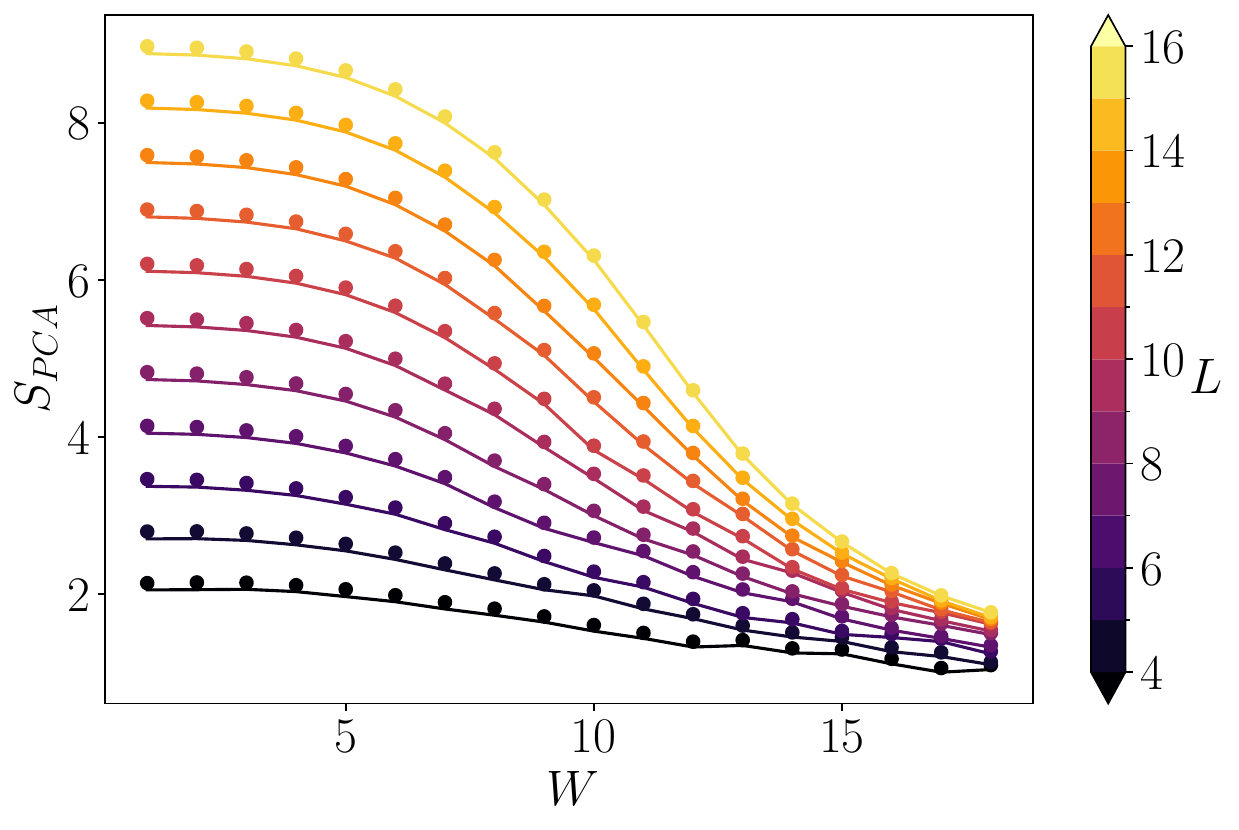}  
    \caption{Comparison of $S_{\mathrm{PCA}}$ (points) without centering the data and participation entropy (lines) for the `N encoding' case. $N_r = 5 \mathcal{N}$ samples have been used.}
    \label{fig:NencodingSPCA}
\end{figure}
In Fig.~\ref{fig:NencodingSPCA} we employ the `$N$-encoding' described before sampling $N_r \sim 5 \, \mathcal{N}$ configurations for each eigenstate. We show a plot of $S_{\textrm{PCA}}$ as a function $W$ (disorder strength) for different sizes of the graph. The points correspond to $S_{\textrm{PCA}}$ while the lines are the estimated participation entropies for different $L$, averaged over $\sim O(10^3)$ realizations of disorder. We observe that they coincide as suggested by Eq.~\eqref{eq:SPCAparticipationentropy}. The interest in the participation entropy relies on the fact that from it one can obtain the fractal dimension of the wavefunction via the definition
\begin{equation}
    D \equiv \frac{\partial S}{\partial \ln N} = \frac{1}{\ln 2}\frac{\partial S}{\partial L}. 
\end{equation}
Many works have addressed the properties of the fractal dimension (see e.g.~\cite{sierant2023universality,vanoni2023renormalization,kravtsov2018non,kutlin2023anatomy,Sierant2022Universal}) as it can be used as an order parameter for the localization transition. 
In fact, if a wavefunction is localized, with localization length $\xi \ll L$, then by increasing the system size no change in $S$ occurs, and thus $D \to 0$. 
If, on the other hand, the wavefunction is delocalized, it will have support over the whole system, and the participation entropy will be $S= L \ln 2$ (assuming $|c_x|^2 = 2^{-L}$). Thus, for $L \to \infty$, $0 \leq D \leq 1$. 

At this point, it is natural to consider the PCA entropy for the more efficient `$L$-encoding', but one does not recover the eigenfunction participation entropy. 
This can be understood from the analogous of Eq.~\eqref{eq:matrix_X} for the `$L$-encoding': now, for each $j$, there will be $k_j$ rows containing a vector whose entries are the binomial representation of the number $j$. 
Of course, such rows can have more than one entry with value $1$, and when taking the product $X^T X$ this will not give simply the number $k_j$, and thus one cannot recover the coefficients $|c_j|^2$. 
Alternatively, this can be understood from the fact that, with the `$L$-encoding', the matrix $C$ has rank $L = \log_2 \mathcal{N}$, while in general there are $\mathcal{N}$ non-zero wavefunction coefficients, meaning that the eigenvalues of $C$ will be a non-trivial combination of the $|c_j|^2$.
Consequently, the form of the PCA entropy for the `$L$-encoding' will be different from the one of the `$N$-encoding', as shown in Fig.~\ref{fig:SPCA_L} $(i)$. 
Naturally, the decrease with $W$ is present also for the `$L$-encoding', reflecting the fact that fewer configurations are sampled with high probability at large $W$. However, the quantitative behavior is different, and no easy way of obtaining information about the transition point has been found. 
For example, by rescaling SPCA with $\ln L$ (see Fig.~\ref{fig:SPCA_L} $(ii)$), i.e. the value in the ergodic phase where all eigenvalues are equal, one gets crossing points between curves for different sizes. Despite being roughly at the correct value of $W$, the crossing points move to smaller $W$ when increasing $L$, which is the opposite behavior with respect to the expected one --- namely with respect to what happens, for instance, in the case of the participation entropy.

\subsubsection{Extrapolation of $W_c$}

\begin{figure}
    \centering
    \includegraphics[width=0.5\linewidth]{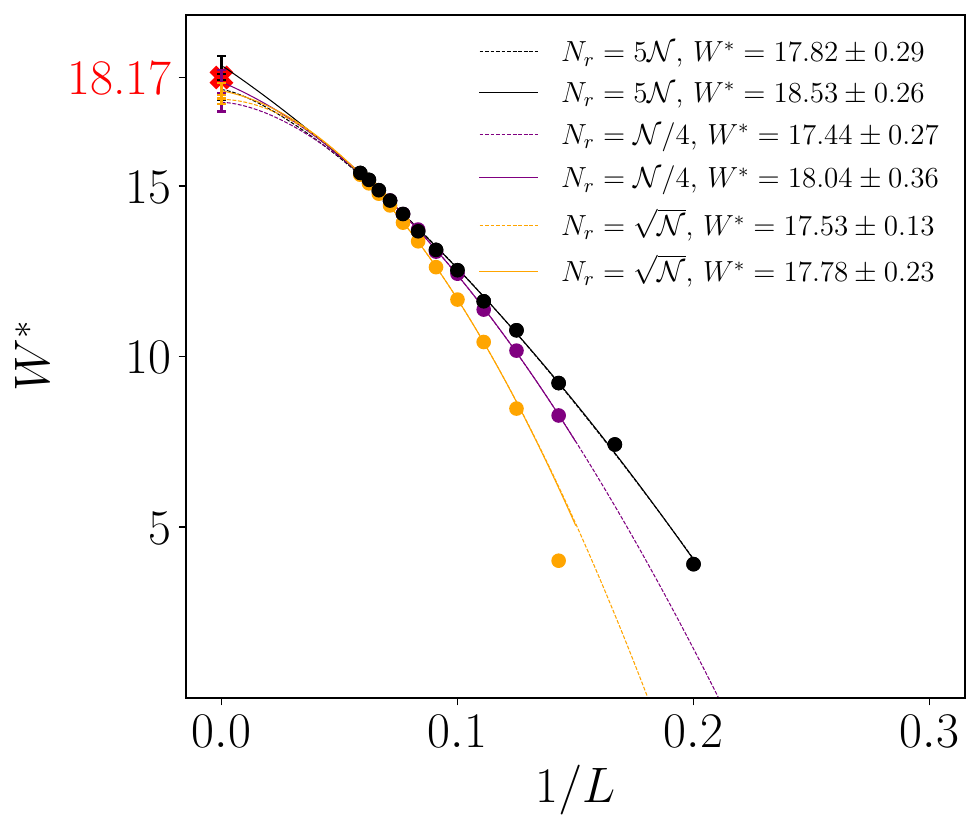}
    \caption{Extrapolation of the critical disorder, as shown in Fig. (2 - ii) of the main text. The solid lines are fits of the form $W^* = a + b/L + c/L^2$ (the same shown in the main text), while the dashed lines are fits of the form $W^* = a + b L^{-c}$}
    \label{fig:w1logw1_fit}
\end{figure}

In this Section, we briefly comment on the extrapolation of $W_c$ shown in the main text. As discussed in the main text, the quadratic fit in $1/L$ gives results perfectly compatible with the known results from different methods. Different fits for the same data are possible, and taking inspiration from the analysis of crossing points in the $r$-parameter shown in Ref.~\cite{sierant2023universality}, we show in Fig.~\ref{fig:w1logw1_fit} also the results for the fit $W^* = a + b L^{-c}$ and the corresponding extrapolated values of $W_c$.

While giving qualitative good results, for some data sets the fit $W^* \sim L^{-c}$ gives less precise results, and overall it is less stable upon addition or removal of fitting points. This suggests that for having a more precise extrapolation value using this fit, data at larger sizes are needed, and this goes beyond the scope of this work.

\section{Study of many-body localization}

In the main text, we have shown how the method presented in this work, based on the principal-component analysis of the wave-functions samplings, allows us to predict with good accuracy the critical properties of the Anderson transition on RRGs. As already mentioned, the choice of that model is twofold: on the one hand, its critical properties are known from other methods, allowing us to benchmark the predictive power of our method. On the other hand, despite being a single-particle problem, it displays features that are typical of interacting systems, making it a non-trivial model to study.

\begin{figure}
    \centering
    \includegraphics[width=\textwidth]{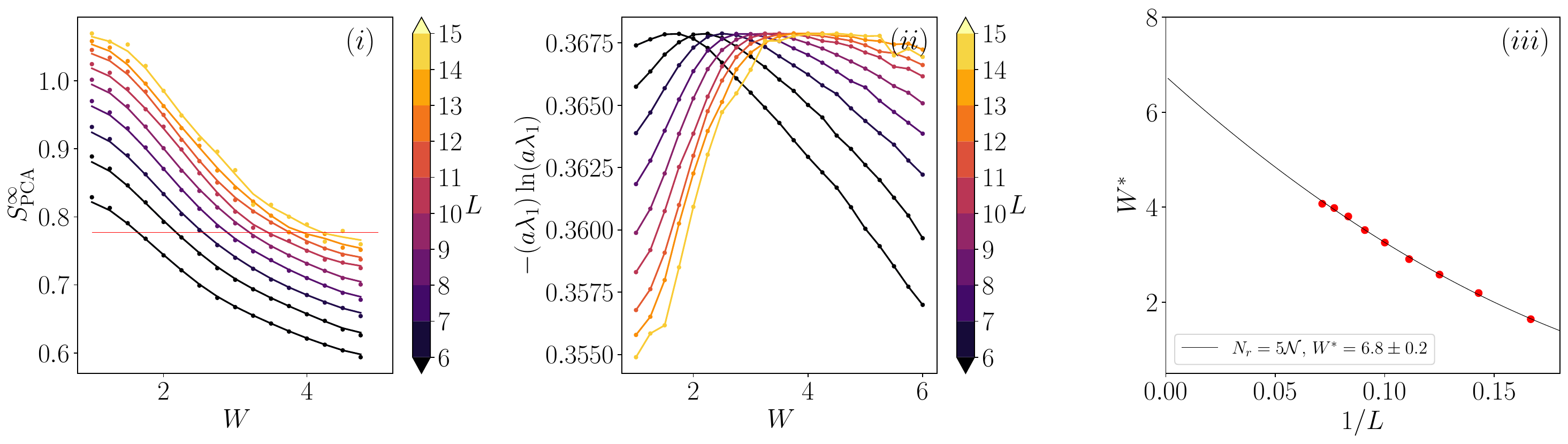}
    \caption{$(i)$ Plot of $S_{\mathrm{PCA}}^{\infty}$ for the ‘Imbrie model' defined in Eq.~\eqref{eq:Imbrie_model}.We look at the crossing point between $S_{\mathrm{PCA}}^{\infty}$ and the horizontal line in red to identify the critical point. With the available sizes, the optimal choice is the line $S_{\mathrm{PCA}}^{\infty}(W^*(L),L) = 1 + \ln a$, with $a=0.8$. The values $W^*(L)$ of the crossing points correspond to the positions of the maxima of the function $- (a \lambda_1) \ln (a \lambda_1)$. In the plot, the dots are the numerical data, while the continuous lines are obtained by applying a Gaussian filter to the data and are a guide for the eye. 
    %However, we used the bare data for the analysis.
    $(ii)$ Plot of $- (a \lambda_1) \ln (a \lambda_1)$, with $a=0.8$. The position of the maxima is identified by the red points. The points flow with increasing system size towards the critical value of the disorder.
    %as the critical disorder, as we expect all $\lambda_i$'s to be of the same magnitude in the delocalized phase, while $\lambda_1=O(1)$ and $\lambda_{i>1}$ exponentially small in the localized phase.
    $(iii)$ Plot of $W^*(L)$ in $1/L$ scale. For the model investigated the slope of the points $(W^*(L),1/L)$ grows with $L$. A quadratic fit gives a good interpolation of the points and predicts a critical value of the disorder $W_c = 7.0 \pm 0.3$. This result is in agreement with previous results in the literature~\cite{Abanin2021distinguishing}.
    }
    \label{fig:MBL_SPCA}
\end{figure}

In this Section, we want to show explicitly that our method can be applied, without modifications, to genuine disordered interacting systems that are believed to present a localization transition. For this purpose, we consider the ‘Imbrie model'~\cite{Imbrie2016manybody,Abanin2021distinguishing}, defined by the Hamiltonian
\begin{equation}
\label{eq:Imbrie_model}
    H = \sum_{i=1}^{L-1}J_i \sigma_i^z \sigma_{i+1}^z + \sum_{i=1}^L \left( h_i \sigma_i^z + \sigma_i^x \right),
\end{equation}
where $\sigma_i^{\alpha}$ ($\alpha = x, \, y, \, z$) are the Pauli matrices on site $i$, $J_i \in [0.8,1.2]$ and $h_i \in [-W,W]$. 
This model has been used in Ref.~\cite{Imbrie2016manybody} to prove the existence of many-body localization in quantum spin chains at infinite temperature, despite there are concerns about the validity of the proof~\cite{Morningstar2022Avalanches,Luitz2017How,DeRoeck2017Stability}. 
%We chose the ‘Imbrie model' rather than the “standard model of MBL", i.e. the XXZ model with random fields, as it will allow us to use our encoding procedure directly, as it does not conserve the total $S_z$ magnetization. 
%In fact, in the presence of total magnetization conservation, one needs to restrict to a symmetry sector, typically $S_z = 0$. 
%This procedure, however, leads to a smaller Hilbert space dimension and a smaller basis set, and, consequently, a different encoding procedure is needed.
The results are displayed in Fig.~\ref{fig:MBL_SPCA}.
%################################################################
%################################################################
%################################################################
%################################################################

The analysis performed is exactly the same we have used in the main text for the RRG. We consider the crossing point of $S_{\mathrm{PCA}}^{\infty}$ with a horizontal line and analyze the flow of the crossing points $W^*(L)$ when the system size is increased, as we show in Fig.~\ref{fig:MBL_SPCA} $(iii)$. Our analysis gives a finite value of the critical disorder $W_c = 7.0 \pm 0.3$, compatible with the literature~\cite{Abanin2021distinguishing}. However, the behavior of $W^*(L)$ might change when the system sizes are increased by orders of magnitude, leading to a larger value of $W_c$. Speculating on the true position of the localization transition is beyond the scope of this work. The inability to perform quantitative comparisons is the main motivation that has led us to use the Anderson model on RRGs as the benchmark for our method.

\end{document}